\documentclass[conference]{IEEEtran}
\IEEEoverridecommandlockouts
\usepackage{makecell}
\usepackage{cite}
\usepackage{todonotes}
\usepackage{amsmath,amssymb,amsfonts}

\usepackage{algorithmic}
\usepackage{graphicx}
\usepackage{textcomp}
\usepackage{xcolor}
\newtheorem{theorem}{Theorem}

\newtheorem{remark}{Remark}
\usepackage{bbm}

\usepackage{bm}

\usepackage{setspace} 
\usepackage[letterpaper, top=0.75in, bottom=1.1in, left=0.67in, right=0.67in]{geometry}
\begin{document}
\setlength{\abovedisplayskip}{0.085cm}
\setlength{\belowdisplayskip}{0.08cm}

\title{Maximum Channel Coding Rate of Finite Block Length MIMO Faster-Than-Nyquist Signaling
\thanks{This work was funded in part by a
Discovery Grant awarded by the Natural Sciences and Engineering
Research Council of Canada (NSERC), 
 in part by the Scientific and Technological Research Council of Turkey, TUBITAK, under grant 122E248.}}

\author{Zichao~Zhang\textsuperscript{\P},
		~Melda~Yuksel\textsuperscript{\S},~Halim~Yanikomeroglu\textsuperscript{\P},~Benjamin~K.~Ng\textsuperscript{\dag},~Chan-Tong~Lam\textsuperscript{\dag}  \\
  \textsuperscript{\P}Department of  Systems and Computer Engineering, Carleton University, Ottawa, ON, Canada  \\
  \textsuperscript{\S}Department of Electrical and Electronics Engineering, Middle East Technical University, Ankara, Turkey  \\
  \textsuperscript{\dag}Faculty of Applied Sciences, Macao Polytechnic
 University, Macao SAR, China \\ \\
 Emails: zichaozhang@cmail.carleton.ca, ymelda@metu.edu.tr, halim@sce.carleton.ca, \{bng, ctlam\}@mpu.edu.mo 
}

\maketitle

\begin{abstract}
The pursuit of higher data rates and efficient spectrum utilization in modern communication technologies necessitates novel solutions. In order to provide insights into improving spectral efficiency and reducing latency, this study investigates the maximum channel coding rate (MCCR) of finite block length (FBL) multiple-input multiple-output faster-than-Nyquist (FTN) channels. By optimizing power allocation, we derive the system's MCCR expression. Simulation results are compared with the existing literature to reveal the benefits of FTN in FBL transmission. 
\end{abstract}

\section{Introduction}
 The ever-growing need for higher data rates and more efficient spectrum utilization poses challenges to the development of the new generation of communication technologies. The novel usage scenario for  6G and beyond such as ultra-reliable low-latency communications (URLLC) requires extremely low latency, which is required to be ten times smaller than the one in LTE standards,  with the guarantee of quality of service (QoS) \cite{5G}. In the classic literature of information theory, the channel capacity is achieved when the block length goes to infinity. {However, longer block length results in longer transmission and longer processing times, making it impractical for URLLC applications.} The study of finite block length (FBL) information theory provides an indication that it is possible to do transmission with a specified reliability at a finite block length. Thus, non-asymptotic information theory is a promising solution to some of the use cases in 6G and beyond, such as internet of
things and machine-to-machine communications \cite{urllc}.  

Multiple-input multiple-output (MIMO) systems have emerged as a pivotal technology, offering significant improvements in channel capacity compared to single-input single-output (SISO) systems \cite{MIMO}. { In MIMO, capacity increases linearly with the minimum of the number of transmit and receive antennas and thus offers spatial multiplexing gain}. MIMO has undergone further advancements, evolving into massive MIMO \cite{massivemimo} and cell-free MIMO \cite{ngo2017cell}.
The study of MIMO capacity typically involves capacity analysis using Shannon's theory, which assumes an infinite block length. However, this assumption is unrealistic in practical applications, making the consideration of FBL MIMO scenarios essential for real-world settings. 
In \cite{fblmimolitrev}, the authors studied the required number of transmit
and/or receive antennas satisfying various error probability/throughput requirements in FBL MIMO systems. In \cite{fblmmimo}, the authors  investigated the expectation and variance of the maximal
achievable rate of FBL massive
MIMO, 
 and the closed-form expressions for expectation and variance of channel dispersion in a massive MIMO system were obtained.

 
 The faster-than-Nyquist (FTN) signaling concept, as an extension of Nyquist's theorem, improves spectral efficiency by transmitting data symbols at a rate that surpasses the Nyquist rate while introducing inter-symbol interference (ISI). Ever since the pioneering work of Mazo \cite{mazo} came out, there have been numerous works on different aspects of FTN, such as information-theoretic studies, signal detection problems, and pulse shape design \cite{takumisurvey}. FTN improves spectral efficiency without requiring more transmission power, making it favorable for energy-efficient communication and a viable candidate for 6G and beyond \cite{amacftn,Zhang2023MIMOAM}. The integration of MIMO with FTN signaling has been shown to be promising in \cite{mimoftn}. {Therefore, it is reasonable to study the performance of MIMO FBL FTN systems, as both MIMO and FTN improve spectral efficiency, while FBL communication lowers latency. }
 
 Polyanskiy \textit{et al.} derived rigorous non-asymptotic achievability and converse bounds for FBL transmission. They also extended the study to other types of channels such as parallel AWGN channels \cite{thesis}. In \cite{schober}, based on the results of \cite{polyangski} and \cite{Tomaso}, the authors derived the maximum channel coding rate (MCCR) of the SISO FBL FTN system.
However, their input power constraint is not the one at the output of the FTN modulation, but at the input, and therefore the result does not represent the exact MCCR. {Concurrently and independently from our work, Kim \cite{kim2023merits} also examined the MCCR of SISO FTN. However, \cite{kim2023merits} assumes a fixed time-bandwidth product.}

In this paper,  we perform a rigorous exploration of the MCCR of FBL MIMO FTN channels {for a fixed number of symbols}. In Section \ref{sec:model}, we derive the system model for FBL MIMO FTN system, in Section \ref{sec:capderi}, we form the MCCR problem. We then use optimal power allocation to solve it and get the final expression. In Section \ref{sec:simul} we provide the simulation results. Finally, in Section \ref{sec:conclu} we conclude the paper. 
\vspace{-5mm}

\section{System Model}
\label{sec:model}
\vspace{-2mm}

Assume the transmitter has $K$ antennas and the receiver has $M$ antennas. The input alphabet is $\mathcal{A}\in\mathbb{R}^{N\times K}$, where $N$ is the number of symbols we transmit.  The transmitted signal from the $k$th antenna is written as 
\begin{equation}
    x_k(t)=\sum_{n=0}^{N-1}a_k[n]p(t-n\delta T), \label{eqn:x(t)}
\end{equation}
where $a_k[n]$ is the $n$th transmitted symbol from the {$k$th antenna, {$k=0,\dots,K-1$}. The parameter $T$ is the signaling period and $\delta\in(0,1]$ is the acceleration factor. Note that in \eqref{eqn:x(t)} there is intentional ISI, as the effective signaling interval is $\delta T$.} We modulate the symbols using a pulse-shaping filter with the impulse response $p(t)$.
Since $p(t)$ is band-limited with bandwidth $\frac{1}{T}$, we can represent it with a set of orthonormal basis signals \cite{Zayed1993AdvancesIS} as
\begin{equation}
    p(t)=\sum_{l=-\infty}^{\infty}p_l\phi(t-l\delta T),
\end{equation} where
\begin{equation}
    \int_{-\infty}^{\infty}\phi(t-i\delta T)\phi^*(t-j\delta T)dt=\left\{\begin{aligned}
        0, i\neq j \\ 1, i=j
    \end{aligned}\right. ,\label{eqn:simplification}
\end{equation} for $i,j \in \mathbb{Z}$, and $^*$ denotes the conjugate operation. The coefficient $p_l,  l\in\mathbb{Z}$ is the projection of $p(t)$ on $\phi(t-l\delta T)$. In other words \begin{equation} p_l=\int_{-\infty}^{+\infty}p(t)\phi(t-l\delta T)dt.\end{equation} The Fourier transform of $\phi(t)$ needs to be constant over the bandwidth $\frac{1}{\delta T}$ {so that $\delta T$ orthogonality in \eqref{eqn:simplification} holds.} 
Therefore, \eqref{eqn:x(t)} can be written as
\begin{equation}
    x_k(t)=\sum_{n=0}^{N-1}\sum_{l=-\infty}^{\infty}a_k[n]p_l\phi(t-(l+n)\delta T).
\end{equation}

The MIMO channel experiences quasi-static fading, and we denote the channel coefficient between the $k$th transmit antenna, {$k= 0,\ldots,K-1$}, and the $m$th receive antenna, {$m=0, \dots, M-1$}, as $h_{mk}$. {We then define the channel matrix as $\bm{H}$, which is an $M\times K$ matrix with entries $(\bm{H})_{m,k}=h_{mk}$.} Then, the received signal at the $m$th receive antenna, {$m=0,\ldots,M-1$}, becomes
{\begin{align}
\tilde{y}_m(t) = \sum_{k=0}^{K-1}h_{mk}x_k(t)+n_m(t),
\end{align}}where $n_m(t)$ is the zero mean, circularly symmetric complex Gaussian noise at the $m$th receive antenna. We assume the noise variance is $\sigma_0^2=\mathbb{E}[n(t)n^*(t)]$, where $\mathbb{E}$ is the expectation. The received signal at each of the antennas is then correlated with a set of receiver filters $\phi^*(t-j\delta T)$. 
Therefore, we can write the $j$th sample, $j \in \mathbb{Z}$, at the output of the receiver filter, $y_m[j]$, as 
{\begin{align}
    y_m[j]&= \int_{-\infty}^{\infty}\tilde{y}_m(t)\phi^*(t-j\delta T)dt\\
&=\sum_{k=0}^{K-1}h_{mk}\sum_{n=0}^{N-1}a_k[n]p_{j-n}+\eta_m[j], \label{eqn:inputoutput}
\end{align}}where $\eta_m[j]=\int_{-\infty}^{\infty}n(t)\phi^*(t-j\delta T)dt$ is the noise sample at the $m$th receive antenna, and \eqref{eqn:inputoutput} is because of the orthogonality defined in \eqref{eqn:simplification}. 

In order to have sufficient statistics, we need to keep all the $y_m[j]$ samples for all $j \in \mathbb{Z}$ \cite{schober}. However, in practice, $p_l$ will become negligible when $l$ is large enough, so we assume that 
$p_l=0$ for $|l|>L$. In other words, $L$ is the time truncation parameter after which $p_l$ becomes negligible. Thus, we only need $N+2L$ samples to detect $N$ samples. For this finite case, we can also write \eqref{eqn:inputoutput} in a matrix multiplication form as 
{\begin{equation}
    \bm{y}_m=\sum_{k=0}^{K-1}h_{mk}\bm{P}\bm{a}_k+\bm{\eta}_m,
\end{equation}}
where  
{\begin{align}
\bm{y}_m &=\left[y_m[-L],\dots,y_m[N+L-1]\right]^T, \\
\bm{\eta}_m & = \left[\eta_m[-L],\dots,\eta_m[N+L-1]\right]^T, \\
\bm{a}_k&=\left[{a}_k[0],\dots,{a}_k[N-1]\right]^T,
 \end{align}}and $^T$ is the transpose operator. The structure of the $(N+2L)\times N$ matrix $\bm{P}$ is given as {$(\bm{P})_{i,j}=p_{i-j-L}, 1\leq i\leq N+2L, 1\leq j\leq N$. We let $\bm{p}_n, n=0,\dots, N-1$ to be the $n$th column of the matrix $\bm{P}$, and we write $\bm{P}=[\bm{p}_0,\dots,\bm{p}_{N-1}]$.} The colored noise $\bm{\eta}_m$ is uncorrelated with the covariance matrix 
\begin{equation}
\mathbb{E}\left[\bm{\eta}_m\bm{\eta}_m^\dagger\right]=\sigma_0^2\bm{I}_{N+2L},
\end{equation}
where $\dagger$ is the Hermitian transpose,  and $\bm{I}_{N+2L}$ is the identity matrix of size $(N+2L)\times(N+2L)$. The input-output relationship of the channel can therefore be written as 
\begin{equation}    \bm{Y}=\left(\bm{H}\otimes\bm{P}\right)\bm{A}+\bm{\Omega},\label{eqn:matchnlmodel}
\end{equation}
where $\otimes$ is the Kronecker product, {$\bm{Y}=\left[\bm{y}_0^T,\dots,\bm{y}_{M-1}^T\right]^T$, $\bm{A}=\left[\bm{a}_0^T,\dots,\bm{a}_{K-1}^T\right]^T$, and $\bm{\Omega}=\left[\bm{\eta}_0^T,\dots,\bm{\eta}_{M-1}^T\right]^T$}.

\section{Capacity Derivation }
\label{sec:capderi}

In this section, we will decompose the FBL MIMO FTN channel into a series of parallel channels and then derive the power constraint expression. Based on that, we optimize the input power distribution and obtain the MCCR 
for FBL MIMO FTN channel. {The proofs and derivations in this section are special for MCCR.}

\begin{theorem}{\cite[Theorem 78]{thesis}} \label{thm:fbl} 
    The maximum channel coding rate of the finite block length parallel AWGN channel is a function of both the block length $N$ and the error probability $\epsilon$, and is given as 
\begin{equation}
    C(N,\epsilon)=C_K-\sqrt{\frac{V_K}{N}}Q^{-1}(\epsilon)+\frac{\log_2(N)}{2N}+\mathcal{O}\left(\frac{1}{N}\right), \label{eqn:fbldef}
\end{equation}
where $C_K$ is the capacity of $K$ parallel AWGN channels and $V_K$ is the channel dispersion. The definition of $C_K$ and $V_K$ can be found in \cite[(4.229), (4.230)]{thesis}.
\end{theorem}  

In MIMO FTN, we can also express the channel as composed of $KN$ parallel {complex Gaussian}  channels, which we prove next.

Assume that the matrices $\bm{H}$ and $\bm{P}$ have the respective singular value decomposition expressions
\begin{align}
    \bm{H}&=\bm{U_H\Sigma_HV_H}^\dagger, \\
    \bm{P}&=\bm{U_P\Sigma_PV_P}^\dagger, \label{eqn:Pdecompose}
\end{align}
where $\bm{U_H}$, $\bm{U_P}$, $\bm{V_H}$ and $\bm{V_P}$ are orthonormal matrices with size $M\times M$, $(N+2L)\times(N+2L)$, $K\times K$ and  $N\times N$, respectively. The $M \times K$ matrix $\bm{\Sigma_H}$ has the entries $\left[\sigma_h[0],\dots,\sigma_h[\text{min}(K,M)-1]\right]$ on its main diagonal. Similarly, the main diagonal for the $(N+2L)\times N$ matrix $\bm{\Sigma_P}$ is $\left[\sigma_p[0],\dots,\sigma_p[N-1]\right]$.
{We define $D=\min\{K, M\}$ so that our derivation is general for any MIMO size.}
By applying the mixed product property of the Kronecker product \cite{matrix}, we can decompose the matrix $\bm{H}\otimes\bm{P}$ as 
\begin{equation}
    \bm{H}\otimes\bm{P}=(\bm{U_H}\otimes\bm{U_P})\left(\bm{\Sigma_H}\otimes\bm{\Sigma_P}\right)(\bm{V_H}^\dagger\otimes\bm{V_P}^\dagger).
\end{equation} 
Then, we multiply the left-hand side of  \eqref{eqn:matchnlmodel} by $(\bm{U_H}^\dagger\otimes\bm{U_P}^\dagger)$ to get 
\begin{eqnarray} (\bm{U_H}^\dagger\otimes\bm{U_P}^\dagger)\bm{Y}  &=&\left(\bm{\Sigma_H}\otimes\bm{\Sigma_P}\right)(\bm{V_H}^\dagger\otimes\bm{V_P}^\dagger)\bm{A} \notag\\
   &&{+}\:(\bm{U_H}^\dagger\otimes\bm{U_P}^\dagger)\bm{\Omega}. \label{eqn:decompchnl}
\end{eqnarray}
Defining 
\begin{align}
\bar{\bm{Y}}&\triangleq(\bm{U_H}^\dagger\otimes\bm{U_P}^\dagger)\bm{Y},\\ \bar{\bm{A}}  &= (\bm{V_H}^\dagger\otimes\bm{V_P}^\dagger)\bm{A},\\ \bar{\bm{\Omega}}&\triangleq(\bm{U_H}^\dagger\otimes\bm{U_P}^\dagger)\bm{\Omega},
\end{align}
note that $\bar{\bm{A}} \in \mathbb{C}^{KN\times 1}$, $\bar{\bm{Y}}\in\mathbb{C}^{M(N+2L)\times 1}$ and $\bar{\bm{\Omega}}\in\mathbb{C}^{M(N+2L)\times 1}$. We have
\begin{align}
    \bar{\bm{A}}&=[\bar{a}_0[0],\dots,\bar{a}_0[N-1],\dots,\\
    &\quad\quad\quad\quad\quad\quad\bar{a}_{K-1}[0],\dots,\bar{a}_{K-1}[N-1]]^T, \\
    \bar{\bm{Y}}&=[\bar{y}_0[-L],\dots, \bar{y}_0[N+L-1],  \dots, \notag\\
    &\quad\quad\quad\quad\quad\quad\bar{y}_{M-1}[-L], \dots, \bar{y}_{M-1}[N+L-1]]^T, \\
    \bar{\bm{\Omega}}&=[\bar{\eta}_0[-L],\dots, \bar{\eta}_0[N+L-1],  \dots, \notag \\
    &\quad\quad\quad\quad\quad\quad\bar{\eta}_{M-1}[-L],\dots, \bar{\eta}_{M-1}[N+L-1]]^T.
\end{align}
We can further simplify the input-output relationship in \eqref{eqn:decompchnl} as
\begin{equation}  \bar{\bm{Y}}=\left(\bm{\Sigma_H}\otimes\bm{\Sigma_P}\right)\bar{\bm{A}}+\bar{\bm{\Omega}}.\label{eqn:barY}
\end{equation}
It is easy to see that the noise component $\bar{\bm{\Omega}}=(\bm{U_H}^\dagger\otimes\bm{U_P}^\dagger)\bm{\Omega}$ is uncorrelated.    
Therefore, we have turned the composite FBL MIMO FTN channel into a collection of $DN$ parallel complex Gaussian channels, where each of the parallel channels has the gain equal to the product {$\sigma_h[d]\sigma_p[n], 0\leq d \leq D-1, 0\leq n \leq N-1$}.    

Next, we start computing the power of the transmitted signal $P_{TX}$ in \eqref{eqn:longstart}-\eqref{eqn:longend} {on the top of the next page.} When we let $\bm{p}_n$ to be the $n$th column of the matrix $\bm{P}$, \eqref{eqn:longend} continues as 
\begin{figure*}
    \begin{align}
        P_{TX}&=\mathbb{E}\left[\frac{1}{N\delta T}\sum_{k=1}^{K}\int_{-\infty}^{\infty}\left|x_k(t)\right|^2dt\right] \label{eqn:longstart}\\
    &=\mathbb{E}\left[\frac{1}{N\delta T}\sum_{k=1}^{K}\int_{-\infty}^{\infty}\left(\sum_{n=0}^{N-1}\sum_{l=-L}^{L}a_k[n]p_l\phi(t-(l+n)\delta T)\right)
    \left(\sum_{m=0}^{N-1}\sum_{s=-L}^{L}a_{k}[m]p_s\phi(t-(s+m)\delta T)\right)^*dt\right] \\
 &=\frac{1}{N\delta T}\sum_{k=1}^{K}\int_{-\infty}^{\infty}
 \sum_{n=0}^{N-1}\sum_{l=-L}^{L}\sum_{m=0}^{N-1}\sum_{s=-L}^{L}\mathbb{E}\left[a_k[n]a_{k}^*[m]\right]p_l\phi(t-(l+n)\delta T)
    p_s^*\phi^*(t-(s+m)\delta T)dt \\
    &=\frac{1}{N\delta T}
 \sum_{n=0}^{N-1}\sum_{m=0}^{N-1}\sum_{k=1}^{K}\mathbb{E}\left[a_k[n]a^*_{k}[m]\right]\sum_{l=-L}^{L}\sum_{s=-L}^{L}p_l
    p_s^*\int_{-\infty}^{\infty}\phi(t-(l+n)\delta T)\phi^*(t-(s+m)\delta T)dt \\
    &=\frac{1}{N\delta T}
 \sum_{n=0}^{N-1}\sum_{m=0}^{N-1}\sum_{k=1}^{K}\mathbb{E}\left[a_k[n]a^*_{k}[m]\right]\sum_{l=-L}^{L}\sum_{s=-L}^{L}p_l
    p_s^*\delta[l+n-(s+m)] \label{eqn:longend}
    \end{align}   
\end{figure*}
\begin{align}
    P_{TX}&=\frac{1}{N\delta T}\sum_{k=1}^{K}
 \sum_{n=0}^{N-1}\sum_{m=0}^{N-1}\mathbb{E}\left[a_k[n]a^*_{k}[m]\right]\left(\bm{p}_m^\dagger\bm{p}_n\right) \\
 &=\frac{1}{N\delta T}\sum_{k=1}^{K}\mathbb{E}\left[\left(\bm{a}_k^\dagger\bm{P}^\dagger\right)\left(\bm{P}\bm{a}_{k}\right)\right] \\
 &=\frac{1}{N\delta T}\sum_{k=1}^{K}\text{tr}\left(\bm{P}\mathbb{E}\left[\bm{a}_k\bm{a}_{k}^\dagger\right]\bm{P}^\dagger\right) \\
 &=\frac{1}{N\delta T}\text{tr}\left(\left(\bm{I}_K\otimes\bm{P}^\dagger\bm{P}\right)\mathbb{E}\left[\bm{A}\bm{A}^\dagger\right]\right)\\
 &=\frac{1}{N\delta T}\text{tr}\left(\left(\bm{I}_K\otimes\bm{P}^\dagger\bm{P}\right)\bm{\Sigma_A}\right).\label{eqn:endofPTX}
\end{align} We assume that this transmission power is limited by $P$; i.e. $P_{TX} \leq P$.
    By applying the decomposition in \eqref{eqn:Pdecompose} to \eqref{eqn:endofPTX}, we obtain
    \begin{align}
        &\frac{1}{N\delta T}\text{tr}\left(\left(\bm{I}_K\otimes\bm{P}^\dagger\bm{P}\right)\bm{\Sigma_A}\right) \notag\\
        &=\frac{1}{N\delta T}\text{tr}\left(\left(\bm{V_H}\otimes\bm{V_P}\right)\left(\bm{I}_K\otimes\bm{\Sigma_P}^\dagger\bm{\Sigma_P}\right)\left(\bm{V_H}^\dagger\otimes\bm{V_P}^\dagger\right)\bm{\Sigma_A}\right) \notag\\
        &=\frac{1}{N\delta T}\text{tr}\left(\left(\bm{I}_K\otimes\bm{\Sigma_P}^\dagger\bm{\Sigma_P}\right)\right)
        \mathbb{E}\bigg[\left(\left(\bm{V_H}^\dagger\otimes\bm{V_P}^\dagger\right)\bm{A}\right)  \notag \\
        & \quad\quad\quad\quad\quad\quad\quad\quad\quad\quad\quad\quad\quad\times\left(\bm{A}^\dagger\left(\bm{V_H}\otimes\bm{V_P}\right)\right)\bigg]\notag\\
        &=\frac{1}{N\delta T}\text{tr}\left(\left(\bm{I}_K\otimes\bm{\Sigma_P}^\dagger\bm{\Sigma_P}\right)\mathbb{E}\left[\bar{\bm{A}}\bar{\bm{A}}^\dagger\right]\right)\notag \\
        &=\frac{1}{N\delta T}\text{tr}\left(\left(\bm{I}_K\otimes\bm{\Sigma_P}^\dagger\bm{\Sigma_P}\right)\bm{\Sigma}_{\bar{\bm{A}}}\right), \label{eqn:constmat}
    \end{align}

\vspace{-0.1cm}
Before we pose the rate optimization problem, let us write both the input-output relationship in \eqref{eqn:barY} and the power constraint in \eqref{eqn:constmat} in single-letter form. Since  $N+2L>N$, there will be some channels that have zero {gain}, and thus we ignore those channels. The { remaining} parallel complex Gaussian channels have the input-output relationship as  
 \begin{align}
     &\bar{y}_{\left\lfloor\frac{i}{N}\right\rfloor}\left[i-N\left\lfloor\frac{i}{N}\right\rfloor-L\right]\allowdisplaybreaks \notag \\ 
     &=\sigma_h\left[\left\lfloor\frac{i}{N}\right\rfloor\right]\sigma_p\left[i-N\left\lfloor\frac{i}{N}\right\rfloor\right]\bar{a}_{\left\lfloor\frac{i}{N}\right\rfloor}\left[i-N\left\lfloor\frac{i}{N}\right\rfloor\right] \allowdisplaybreaks \notag\\
     &+\bar{\eta}_{\left\lfloor\frac{i}{N}\right\rfloor}\left[i-N\left\lfloor\frac{i}{N}\right\rfloor-L\right], ~~i=0,\dots,{D}N-1,
 \end{align}
 where $\lfloor\cdot\rfloor$ means the floor operation. Also,
 we denote $\sigma_a^2[i]$, $i=0,\dots,{D}N-1$ be the $i$th diagonal value of {$KN\times KN$ matrix} $\bm{\Sigma}_{\bar{\bm{A}}}$.  Notice that $\sigma_a^2[i]$ is also the input power for the $i$th channel, since $\sigma_a^2[i]=\mathbb{E}\left[\left|\bar{a}_{\lfloor\frac{i}{N}\rfloor}[i]\right|^2\right]$. 
 From \eqref{eqn:constmat}, we obtain the power constraint in single letter form by directly multiplying the diagonal values of the diagonal matrices $\bm{\Sigma_P}$ and $\bm{\Sigma}_{\bar{\bm{A}}}$ as 
 \begin{equation}
     \frac{1}{N\delta T}\sum_{i=0}^{{D}N-1}\left|\sigma_p\left[i-N\lfloor\frac{i}{N}\rfloor\right]\right|^2\sigma_a^2[i]\leq P. \label{eqn:powconst}
 \end{equation}
 As a result, we write the optimization problem for computing the capacity for ${D}N$ parallel complex Gaussian channels as 
    \begin{align}
     &C_{{D}N}=\notag\\
     &\underset{\sigma_a^2[i], \forall i}{\max}\sum_{i=0}^{{D}N-1}\log_2\left(1+\frac{\left|\sigma_h[\lfloor\frac{i}{N}\rfloor]\sigma_p[i-N\lfloor\frac{i}{N}\rfloor]\right|^2\sigma_a^2[i]}{\sigma_0^2}\right),\notag
\end{align}
subject to the power constraint \eqref{eqn:powconst}. 
 The Karush-Kuhn-Tucker conditions to solve this problem are  
\begin{subequations}
    \begin{align}
     \Bigg(-\frac{\sigma_0^2\left|\sigma_h[\lfloor\frac{i}{N}\rfloor]\sigma_p[i-N\lfloor\frac{i}{N}\rfloor]\right|^2}{\sigma_0^2+\left|\sigma_h[\lfloor\frac{i}{N}\rfloor]\sigma_p[i-N\lfloor\frac{i}{N}\rfloor]\right|^2\sigma_a^2[i]} &+  \notag\\
      \frac{\mu\left(\left|\sigma_p[i-N\lfloor\frac{i}{N}\rfloor]\right|^2\right)}{N\delta T}   
    -\lambda[i]\Bigg) &=0, \forall i \\
      \frac{1}{N\delta T}\sum_{i=0}^{{D}N-1}\left|\sigma_p\left[i-N\left\lfloor\frac{i}{N}\right\rfloor\right]\right|^2\sigma_a^2[i]& \leq P
      \\ \sigma_a^2[i]\lambda[i]&=0, \forall i \\
      \sigma_a^2[i]&\geq 0, \forall i
\end{align}
\end{subequations}
where $\mu$ and $\lambda[i]$ are dual variables. 
The solution for the optimal power allocation is then written as 
\begin{align}
    \bar{\sigma}_a^2[i]=\frac{\sigma_0^2}{\left|\sigma_p[i-N\times\lfloor\frac{i}{N}\rfloor]\right|^2}\left(\frac{N\delta T}{\mu}-\frac{1}{\left|\sigma_h[\lfloor\frac{i}{N}\rfloor]\right|^2}\right)^+, \label{eqn:wfsolu}
\end{align}
where $(a)^+=\max(a,0)$. The variable $\mu$ can be solved by 
\begin{align}
    \frac{\sigma_0^2}{N\delta T}\sum_{i=0}^{{D}N-1}\left(\frac{N\delta T}{\mu}-\frac{1}{\left|\sigma_h[\lfloor\frac{i}{N}\rfloor]\right|^2}\right)^+=P.
\end{align}

Overall, we obtain ${D}N$ {parallel complex Gaussian} channels, and we use ${D}(N+2L)$ samples to detect ${D}N$ symbols. {Similar to \cite{polyangski}, we optimize the $C_{DN}$ term} by combining these parameters as well as the optimal power allocation and applying Theorem \ref{thm:fbl}, we get the maximum channel coding rate of FBL MIMO FTN in bits/channel use as given in the next theorem. 
\begin{theorem}
     The maximum channel coding rate 
     of the FBL MIMO FTN channel is a function of both the block length $N$ and the error probability $\epsilon$, and is given as 
\begin{align}
    &C(N,\epsilon)=\frac{N}{N+2L}\left(C_{{D}N}-\sqrt{\frac{V_{{D}N}}{{D}N}}Q^{-1}(\epsilon) \notag \right. \\   &\quad\quad\quad\quad\quad\quad\quad\quad\left.+\frac{\log_2({D}N)}{2{D}N}+\mathcal{O}\left(\frac{1}{{D}N}\right)\right), \label{eqn:fblcapacity}
\end{align}
where 
\begin{align}
    &C_{{D}N}=\frac{1}{N}\times \notag \\
    &\quad\sum_{i=0}^{{D}N-1}\log_2\left(1+\frac{\left|\sigma_h[\lfloor\frac{i}{N}\rfloor]\sigma_p[i-N\lfloor\frac{i}{N}\rfloor]\right|^2\sigma_a^2[i]}{\sigma_0^2}\right),
\end{align}
and 
\begin{align}
    &V_{{D}N} =\frac{(\log_2e)^2}{N}\times \notag\\ 
    &\sum_{i=0}^{{D}N-1}\left(1- \left(1+\frac{\left|\sigma_h[\lfloor\frac{i}{N}\rfloor]\sigma_p[i-N\lfloor\frac{i}{N}\rfloor]\right|^2\sigma_a^2[i]}{\sigma_0^2}\right)^{-2}\right).
\end{align}\label{thm:theorem2}
\end{theorem}

\begin{remark}\label{rem:boildown}
    Note that if we set $K=M=1$,  the above result reduces to the maximum channel coding rate of SISO FTN. Note that the SISO result in \cite{schober} is based on the power constraint at the input of FTN modulation, whereas Theorem~\ref{thm:theorem2} uses the power constraint at the output of the transmitter, taking ISI into account.
\end{remark}\vspace{1mm}
\begin{remark} 
When $N$ goes to infinity, we can get the capacity of the MIMO FTN channel as in \cite{mimoftn}.
\end{remark}\vspace{1mm}
\begin{remark}
We can also obtain the MCCR 
in bit/s/Hz by normalizing the bandwidth to obtain 
\begin{equation}
    \bar{C}(N,\epsilon)=\frac{1}{\delta(1+\beta)}C(N,\epsilon), \quad \mathrm{bits/s/Hz}, \label{eqn:bpshzcap}
\end{equation}
{if $p(t)$ is a raised cosine pulse with roll-off factor $\beta\in[0,1]$. }
\end{remark}

\begin{figure}
    \centering
    \includegraphics[scale=0.57]{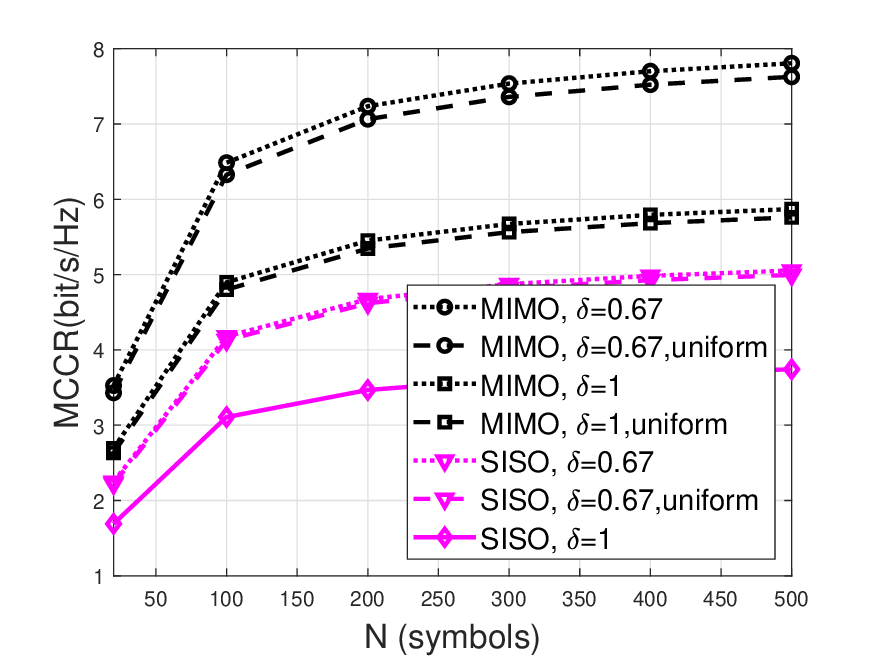}
    \caption{MCCR versus block length $N$ for MIMO and SISO FBL FTN, for $\beta=0.5$, and $\textsc{SNR}=20$ dB. 
    }
    \label{fig:systematicdiffN}
\end{figure}
\section{Simulation Results}
\label{sec:simul}

In this section, we present the performance of FBL MIMO FTN and its improvement over SISO and Nyquist transmission. In the figures, we set the symbol period $T=0.01$, and the number of truncation {$L=10$. For $L=10$ the energy of the pulse $p(t)$ is well contained and the discarded part has energy less than $10^{-4}$.} The MIMO size is $2\times 2$ for all curves. {Raised cosine pulse with roll-off factor $\beta$ is used for $p(t)$ and we use sinc pulse {$\sqrt{\delta T}\sin(\pi t/(\delta T))/\pi t$ for $\phi(t)$}.  All channel coefficients $h_{mk}$ are independent and identically distributed according to the complex Gaussian distribution $\mathcal{CN}(0,1/K)$.} We average all curves over 1000 random channel realizations.

{In Fig. \ref{fig:systematicdiffN}, we set the error probability $\epsilon=10^{-6}$. The signal-to-noise ratio $\textsc{SNR}=\frac{P}{\sigma_0^2}$ is  20 dB. We plot MCCR versus block length $N$ and compare MIMO FBL FTN ($\delta = 0.67$) stated in {Theorem~\ref{thm:theorem2}} with both MIMO FBL Nyquist transmission ($\delta = 1$) and the SISO case ($\delta = 0.67$) to see the performance gains. For $N=500$, FTN increases {Nyquist} MIMO MCCR by approximately 1.93 bits/s/Hz. We also examine the optimal power allocation with uniform power allocation, where the latter is stated as ``uniform'' in the legend to express uniform power allocation in time and/or space. When we compare MIMO FBL FTN ($\delta = 0.67$), which is obtained with the optimal power allocation for MCCR, with uniform power allocation (MIMO, $\delta= 0.67$, ``uniform'') the gain is only 0.18 bits/s/Hz. Therefore, we state that uniform power allocation closely follows the optimal power allocation scenario.
Note that for the SISO, $\delta=1$ case, equal power allocation is inherently optimal. Finally, comparing MIMO curves with SISO curves, either with FTN or for Nyquist, we can easily deduce the gains due to multiple antennas.} 

\begin{figure}[t]
    \centering
    \includegraphics[scale=0.57]{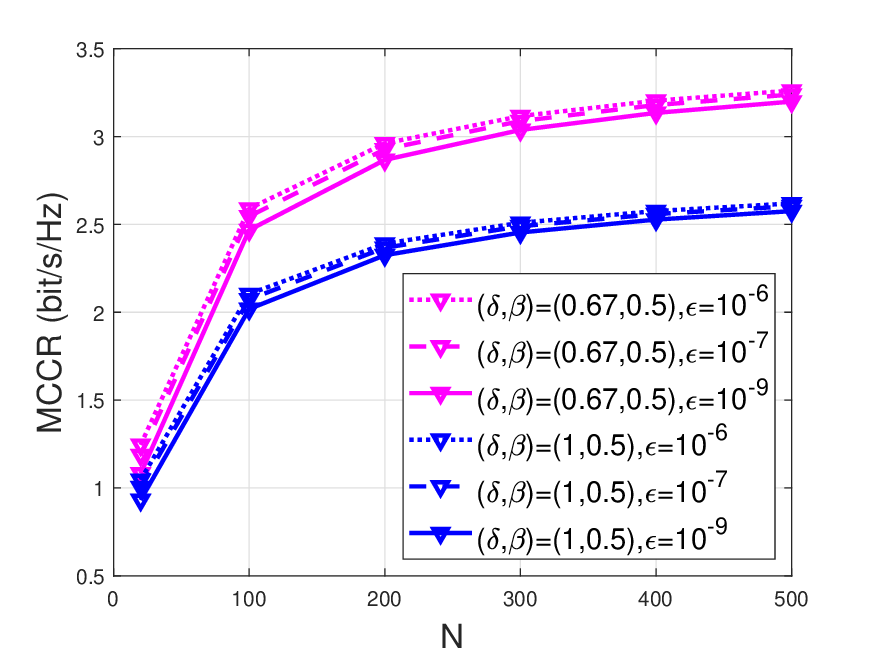}
    \caption{MCCR versus block length $N$ for different $\epsilon$ values, for both FTN ($\delta = 0.67$) and for Nyquist ($\delta = 1$) transmission, for $\beta= 0.5$, and {$\textrm{SNR}= 10$~dB} for $2 \times 2$ MIMO.}
    \label{fig:effectepsidiffN}
\end{figure}
In Fig. \ref{fig:effectepsidiffN}, we display MCCR versus block length $N$ to study the influence of the probability of error, $\epsilon$, for both FTN ($\delta = 0.67$) and for Nyquist ($\delta = 1$) transmission, for $\beta= 0.5$, and {$\textrm{SNR}= 10$ dB} for $2 \times 2$ MIMO.
When we compare the performance for different $\epsilon$ values, we observe that if we pursue a lower probability of error, {for example, if we are going from $\epsilon=10^{-6}$ to $\epsilon=10^{-9}$, the MCCR will only experience 1.93\% decrease when $(\delta, \beta)=(0.67, 0.5)$. Therefore, the price we pay for better performance is affordable}, and FBL MIMO FTN is quite suitable for URLLC applications.

\begin{figure}
    \centering
    \includegraphics[scale=0.6]{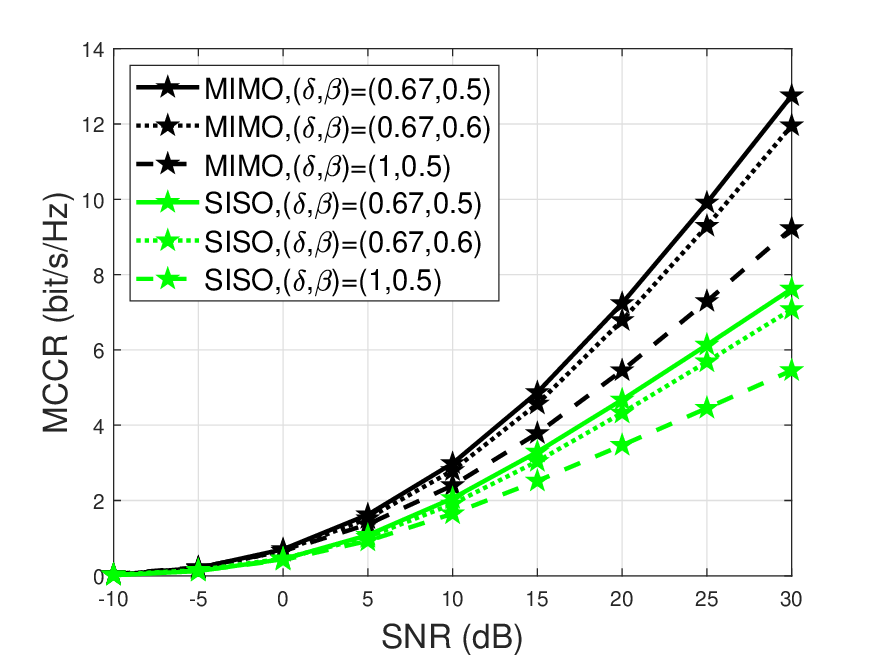}
    \caption{MCCR versus SNR for MIMO and SISO FBL FTN for different $(\delta,\beta)$ pairs, $\epsilon = 10^{-6}$, and $N=100$.}
    \label{fig:systematicdiffSNR}
\end{figure}

In Fig. \ref{fig:systematicdiffSNR}, we set the error probability $\epsilon=10^{-6}$ and $N= 100$. We display MIMO FBL FTN with SISO FBL FTN MCCR as a function of SNR and see the improvement brought by MIMO. This improvement also increases with increasing SNR. We also compare FTN performance for $\delta = 0.67$ with Nyquist signaling with $\delta = 1$ to show the performance improvement brought by FTN both in MIMO and SISO cases. We observe that there is a significant difference between the slope of FTN curves and the slope of Nyquist signaling. This is due to the degrees-of-freedom (DoF) gain brought by FTN. According to the study \cite{dof}, in MIMO spatial DoF gain is due to transmitting independent symbols through virtual parallel channels. Similarly, in FTN signaling, we are able to decompose channels in both space and time into equivalent parallel channels and thus increase DoF.  
The DoF gain of SISO FTN can be computed as the limit of the ratio of {$C_{DN}$} of FTN over the capacity of Nyquist signaling as $P/\sigma_0^2$ goes to infinity. For {raised cosine pulses with roll-off factor $\beta$ and $\frac{1}{1+\beta}\leq\delta\leq 1$} this ratio becomes
\begin{align}
   r&= \underset{\frac{P}{\sigma_0^2}\rightarrow\infty}{\lim}\frac{\frac{1}{2\delta(1+\beta)}\log(1+\frac{\delta TP}{\sigma_0^2})}{\frac{1}{2(1+\beta)}\log(1+\frac{TP}{\sigma_0^2})} = ~\frac{1}{\delta}.
\end{align}
Therefore, when we compare FTN curves and Nyquist signaling curves in Fig. \ref{fig:systematicdiffSNR}, we can see that the slope of FTN is approximately $\frac{1}{\delta}=1.5$ times the slope of the Nyquist curve {for the SISO case.} As the DoF gain from MIMO is $\min\{K,M\}=2$ \cite{dof}, for MIMO transmission, the overall DoF gain from both FTN and MIMO is $\min\{K,M\}/\delta$, and the slope of MIMO FTN curve is approximately 3 times that of SISO Nyquist transmission.
In Fig. \ref{fig:systematicdiffSNR}, we additionally study the impact of $\beta$ on the MCCR. {We can see that for $\delta=0.67$, $\beta =0.5$ outperforms $\beta = 0.6$ both for SISO and MIMO. This suggests that for a fixed $\delta$ value, we should choose $\beta$ closer to $\frac{1}{\delta}-1$, which is also confirmed in \cite{mimoftn}.}  

\begin{figure}[t]
    \centering
    \includegraphics[scale=0.57]{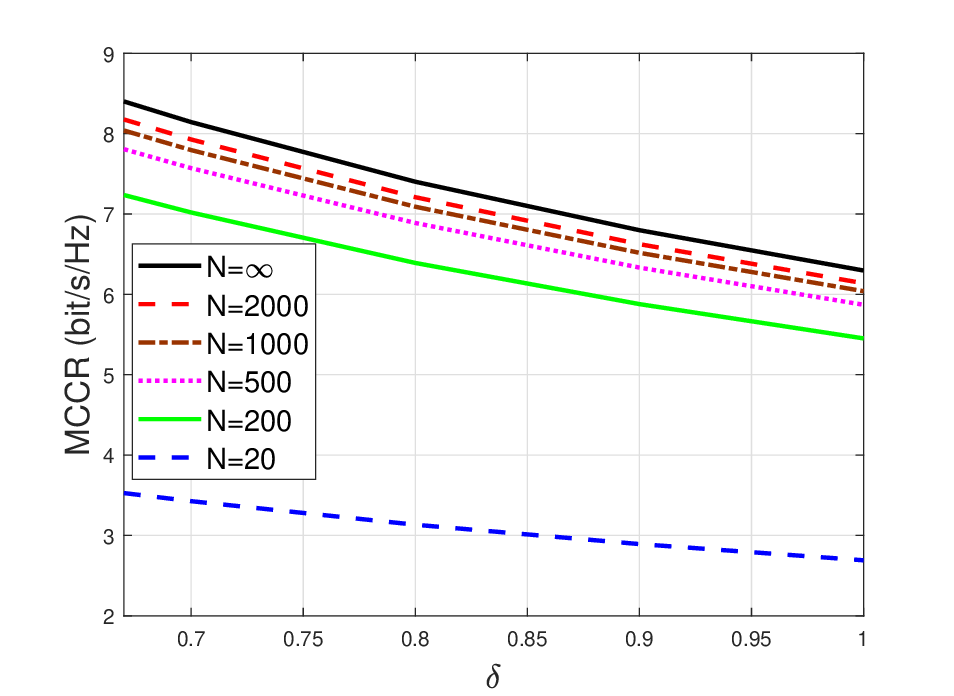}
    \caption{MCCR versus $\delta$ for different $N$ values, with $\text{SNR}=20$ dB, $\epsilon=10^{-6}$, and $\beta=0.5$.} 
    \label{fig:effectepsidiffSNR}
\end{figure}

{In Fig. \ref{fig:effectepsidiffSNR}, we show MCCR versus $\delta$ and also study the influence of $N$ over the MCCR. We also include the infinite block length channel capacity for MIMO FTN \cite{mimoftn} as an upper bound. We can see that the MCCR approaches the channel capacity as the number of symbols increases.  $N = 2000$ attains $97.4\%$ of the capacity, while $N = 200$ attains $86.58\%$ of the capacity. On the other hand, $N = 20$ is severely limited, achieving only $42.73\%$ of the capacity.}

\section{Conclusion}
\label{sec:conclu}

In this paper, we study the MCCR of FBL MIMO FTN systems. By decomposing the MIMO FTN channel into parallel channels, 
we are able to obtain the optimal power allocation for FBL transmission with FTN. The simulation results show the benefits of FTN in FBL MIMO. Our study suggests that FBL MIMO FTN provides significant improvement even for relatively small block lengths, and the gain brought by MIMO and FTN jointly improves spectral efficiency. The ability to achieve a high transmission rate while maintaining exceptional error probability performance positions FBL MIMO FTN as a compelling candidate for URLLC applications.

\bibliographystyle{IEEEtran}
\bibliography{main}

\end{document}